\documentclass[letterpaper,twocolumn,pra,aps,superscriptaddress,amsmath,amssymb,floatfix]{revtex4-1}
\usepackage{amsmath}
\allowdisplaybreaks[4]
\usepackage{amsthm}
\usepackage{amssymb}
\usepackage{color}
\usepackage{graphicx}
\usepackage{epstopdf}
\usepackage{dcolumn}
\usepackage{bm}
\usepackage{bbm}
\usepackage{mathrsfs}
\usepackage{pifont}
\usepackage{float}
\usepackage{appendix}
\usepackage{xcolor} 

\usepackage[colorlinks,linkcolor=blue,anchorcolor=blue,citecolor=blue,urlcolor=blue]{hyperref} 

\newcommand{\bra}[1]{\ensuremath{\left\langle#1\right|}}
\newcommand{\ket}[1]{\ensuremath{\left|#1\right\rangle}}
\newcommand{\braket}[2]{\left< #1  \left| #2 \right. \right>}

\begin{document}
	\title{Topological Phase Transitions and Edge-State Transfer in Time-Multiplexed Quantum Walks}
	\affiliation{Institute of Theoretical Physics and State Key Laboratory of Quantum Optics Technologies and Devices, Shanxi University, Taiyuan 030006, China}
	
	\author{Huimin~Wang}
	\affiliation{Institute of Theoretical Physics and State Key Laboratory of Quantum Optics Technologies and Devices, Shanxi University, Taiyuan 030006, China}
	
	\author{Zhihao Xu}
	\email{xuzhihao@sxu.edu.cn}
	\affiliation{Institute of Theoretical Physics and State Key Laboratory of Quantum Optics Technologies and Devices, Shanxi University, Taiyuan 030006, China}
	
	\author{Zhijian Li}
	\email{zjli@sxu.edu.cn}
	\affiliation{Institute of Theoretical Physics and State Key Laboratory of Quantum Optics Technologies and Devices, Shanxi University, Taiyuan 030006, China}

	
	\begin{abstract}
		
		We investigate the topological phase transitions and edge-state properties of a time-multiplexed nonunitary quantum walk with sublattice symmetry. By constructing a Floquet operator incorporating tunable gain and loss, we systematically analyze both unitary and nonunitary regimes. In the unitary case, the conventional bulk-boundary correspondence (BBC) is preserved, with edge modes localized at opposite boundaries as predicted by topological invariants. In contrast, the nonunitary regime exhibits non-Hermitian skin effects, leading to a breakdown of the conventional BBC. By applying non-Bloch band theory and generalized Brillouin zones, we restore a generalized BBC and reveal a transfer phenomenon, where edge modes with different sublattice symmetries can become localized at the same boundary. Furthermore, we demonstrate that the structure of the spectral loops in the complex quasienergy plane provides a clear signature for these transfer behaviors. Our findings deepen the understanding of nonunitary topological phases and offer valuable insights for the experimental realization and control of edge states in non-Hermitian quantum systems.

	\end{abstract}
	
	\maketitle
	\section{INTRODUCTION}
	Topological phases exhibit remarkable properties that fundamentally challenge our conventional understanding of phases and phase transitions \cite{RevModPhys.83.1057,RevModPhys.82.3045,RevModPhys.88.035005,RevModPhys.88.021004,PhysRevX.9.041015}. Unlike traditional phases, which are characterized by local order parameters, topological phases are defined by nonlocal topological invariants. These invariants dictate the existence and number of topological edge states at interfaces, as described by the bulk-boundary correspondence (BBC) \cite{Ryu_2010,PhysRevB.82.115120,PhysRevB.88.121406,PhysRevLett.110.215301}. Recently, non-Hermitian systems have garnered significant attention due to their unconventional properties, which differ markedly from those of Hermitian systems \cite{PhysRevA.101.013635,PhysRevB.102.035153,PhysRevB.99.201103,PhysRevLett.124.086801,PhysRevX.13.021007,PhysRevB.108.184205,PhysRevLett.116.133903,PhysRevLett.125.226402,PhysRevLett.123.016805,PhysRevLett.121.086803,PhysRevLett.121.136802,PhysRevLett.123.066404,10.1093/ptep/ptaa140,PhysRevLett.121.026808,PhysRevLett.125.126402}. One of the most intriguing phenomena is the breakdown of the conventional BBC in non-Hermitian topological systems \cite{PhysRevLett.116.133903,PhysRevLett.125.226402,PhysRevLett.123.016805,PhysRevLett.121.086803,PhysRevLett.121.136802,PhysRevLett.123.066404,10.1093/ptep/ptaa140,PhysRevLett.121.026808}. In such systems, bulk eigenstates tend to localize near the boundaries due to non-Hermiticity, resulting in the so-called non-Hermitian skin effect (NHSE) \cite{PhysRevLett.121.086803,PhysRevLett.121.026808,PhysRevB.99.201103,PhysRevLett.124.086801,PhysRevX.13.021007,PhysRevLett.121.136802,PhysRevLett.123.066404,10.1093/ptep/ptaa140,PhysRevLett.125.126402}. To accurately capture the topological edge states in non-Hermitian systems, the non-Bloch band theory, based on the concept of the generalized Brillouin zone (GBZ), was developed. This framework extended the standard Bloch band theory, providing a more precise description of topological phenomena in non-Hermitian settings \cite{PhysRevLett.121.086803,PhysRevLett.121.136802,PhysRevLett.123.066404,10.1093/ptep/ptaa140,PhysRevLett.125.186802,PhysRevB.102.035153}. Furthermore, a series of optical experiments were carried out to explore non-Hermitian topological phases, with the associated edge states being directly observed \cite{Poli2015,doi:10.1126/science.aar4005,PhysRevLett.120.113901,PhysRevA.98.063847,doi:10.1126/science.aaz8727,PhysRevLett.115.040402,xiao2017observation,Xiao2020}. These experimental advances offered valuable insights into the behavior of topological edge states in non-Hermitian systems, further deepening our understanding of these exotic phases.
	
	Quantum walks (QWs) \cite{PhysRevA.61.013410,PhysRevLett.106.180403,PhysRevLett.108.010502,PhysRevLett.125.186804,PhysRevA.88.042334,PhysRevA.87.012314,Science2010} are quantum counterparts of classical random walks, describing the coherent propagation of a spin-1/2 particle across a lattice. Unlike their classical analogs, QWs exhibit a characteristic quadratic spreading of the walker, which underpins their unique dynamical behavior. This property has led to extensive applications in quantum computing, search algorithms, quantum simulation, and quantum state transfer. Beyond these applications, QWs serve as a versatile platform for exploring topological phenomena in quantum dynamics  \cite{PhysRevA.48.1687,PhysRevA.82.033429,Science2010,Kitagawa2012,Kitagawa2012j,Cardano2016,PhysRevLett.125.186802,Xiao2020,wang2019observation,PhysRevLett.123.246801,PhysRevLett.126.230402}. A notable advantage of photonic QWs is their ability to incorporate gain and loss, enabling the investigation of topological effects in nonunitary systems \cite{xiao2017observation,Xiao2020,PhysRevLett.115.040402,PhysRevLett.123.246801,PhysRevLett.126.230402,lin2023manipulating,PhysRevA.110.052410,PhysRevLett.119.130501,wang2019observation,PhysRevA.109.012409,mittal2021persistence,meng2022topological,PhysRevA.93.062116}. Importantly, QWs are inherently Floquet systems, where time evolves in discrete steps. This periodic driving gives rise to topological phases that are fundamentally different from those in static Hamiltonians. In such time-periodic systems, the quasienergy is defined modulo $2\pi$, resulting in an unbounded Floquet spectrum. For two-band topological systems, this leads to not only conventional edge states above and below a given Floquet band, but also the possibility of additional edge states crossing the quasienergy Brillouin zone boundary ($E=\pm\pi$) \cite{PhysRevA.82.033429,PhysRevX.3.031005,PhysRevX.4.041048,PhysRevX.4.031027,PhysRevB.88.121406,Kitagawa2012j}. Significant progress has been made in understanding Floquet topological phases in QWs \cite{cardano2017detection,Endo_2017,PhysRevLett.118.130501,10.21468/SciPostPhys.5.2.019,PhysRevLett.121.260501,Mochizuki_2020,PhysRevB.102.035418,Wang:20,PhysRevLett.124.050502,PhysRevLett.127.270602}, with experimental observations providing direct evidence for the existence of both conventional and anomalous topological edge states \cite{PhysRevA.96.033846,Mukherjee2017,PhysRevLett.121.100502,PhysRevA.98.063847,PhysRevLett.126.230402,PhysRevLett.126.230402,10.1093/nsr/nwad005,lin2023manipulating,PhysRevA.110.052410}. 
	Furthermore, the interplay between topology and nonunitarity has revealed a wealth of intriguing physical phenomena, such as the persistence of Floquet topological phases protected by parity and time-reversal symmetry \cite{mittal2021persistence}, and the realization of higher-order non-Hermitian topological phases \cite{meng2022topological}. Notably, Zhang and Gong \cite{PhysRevB.101.045415} demonstrated the existence of anomalous degenerate zero and $\pi$ modes that are simultaneously localized at the same boundary; a scenario distinct from the standard case in one-dimensional (1D) systems, where two degenerate edge modes are typically localized at opposite boundaries. In their study, the transfer behavior of the zero and $\pi$ modes cannot be independently controlled. Whether these anomalous edge modes can be independently manipulated, and the underlying mechanisms governing their behavior, remain pressing open questions \cite{PhysRevB.105.094103,LIAO2024107372}. 
	
	In this work, we investigate a time-multiplexed nonunitary QW with tunable gain and loss. By systematically exploring both unitary and nonunitary regimes, we construct the topological phase diagram and characterize the behavior of edge states. We uncover a transfer phenomenon in which the zero and $\pi$ modes can be independently controlled and localize at the same boundary due to non-Hermitian effects. Furthermore, we demonstrate that the correspondence between spectral features under periodic and fixed boundary conditions provides a clear signature of edge-state localization and transfer. Our results elucidate the influence of nonunitarity on topological edge states in Floquet systems and offer valuable guidance for future experimental and theoretical studies of non-Hermitian QWs.
	
	\section{TIME-MULTIPLEXED NONUNITARY QW}
	
		\begin{figure}
		\centering
		\includegraphics[width=\linewidth]{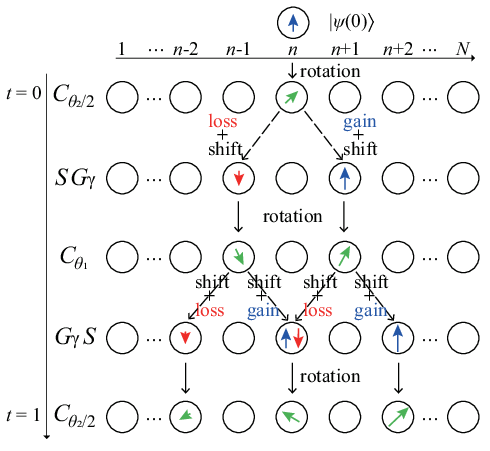}
		\caption{Schematic of a single step in the evolution of a nonunitary QW, initialized in the state $\ket{\psi(t=0)} = \ket{n, H}$. Circles denote lattice sites. Red (blue) arrows indicate the horizontal ($\ket{H}$) [vertical ($\ket{V}$)] polarization states, while green arrows represent superposition states of $\ket{H}$ and $\ket{V}$. The length of each arrow is proportional to the absolute values of the corresponding probability amplitude.}\label{fig0}
	\end{figure} 
	
	We consider a discrete-time nonunitary QW on a 1D finite chain, governed by the Floquet operator \cite{xiao2017observation,PhysRevA.102.062202}
	\begin{equation}
		U=C_{\theta _{2}/2}G_{\gamma} S C_{\theta _{1}} S G_{\gamma} C_{\theta _{2}/2},
		\label{eq1} 
	\end{equation}
	which acts on the tensor product space $\mathscr{H}=\mathscr{H}_{\mathrm{site}} \otimes \mathscr{H}_{\mathrm{coin}}$, where $\mathscr{H}_{\mathrm{site}}$ and $\mathscr{H}_{\mathrm{coin}}$ denote the spatial and coin subspaces, respectively.
	The coin operator $C_\theta$ performs a rotation of the coin state by an angle $\theta$ about the $y$ axis, and is defined as
	\begin{equation}
		C_{\theta}=\sum_{n=1}^{N}|n\rangle\langle n| \otimes \begin{pmatrix} \cos{\theta} & -\sin{\theta} \\ \sin{\theta} & \cos{\theta} \end{pmatrix},
		\label{eq2} 
	\end{equation}
	where $n$ labels the lattice sites and $N$ denotes the system size. The coin-dependent shift operator $S$ moves the walker to neighboring sites depending on its polarized states: horizontal ($\ket{H}$) or vertical ($\ket{V}$). Explicitly, it is given by
	\begin{align}
		S=&\sum_{n=1}^{N-1}\ket{n+1} \bra{n} \otimes \ket{H} \bra{H} +\ket{n} \bra{n+1} \otimes \ket{V} \bra{V}\notag \\ 
		& + P \left(\ket{1} \bra{N} \otimes \ket{H} \bra{H} +\ket{N} \bra{1} \otimes \ket{V} \bra{V}\ \right) \notag \\
		& + (1-P)\left(\ket{1} \bra{1} \otimes \ket{H} \bra{V} + \ket{N} \bra{N}\otimes \ket{V} \bra{H} \right),
		\label{eq3} 
	\end{align}
	where $P=1$ corresponds to periodic boundary conditions (PBCs), and $P=0$ corresponds to fixed boundary conditions (FBCs), in which case the walker's polarization is flipped at the two ends of the chain. Nonunitarity is introduced via the gain and loss operator
	\begin{equation}
		G_\gamma=\sum_{n=1}^{N} \ket{n}\bra{n} \otimes \begin{pmatrix} e^{\gamma} & 0 \\ 0 & e^{-\gamma} \end{pmatrix},
		\label{eq4} 
	\end{equation}
	where $\gamma$ is a tunable gain and loss parameter. The Floquet operator $U$ is applied repeatedly to the walker’s state, governing the discrete-time Floquet dynamics. As illustrated in Fig. \ref{fig0}, the time evolution begins with the application of the coin operator $C_{\theta_2 /2}$, which mixes the polarization states. Subsequently, the gain-loss operator $G_\gamma$ amplifies the $\ket{H}$ component and attenuates the $\ket{V}$ component for $\gamma>0$. The coin-dependent shift operator $S$ then moves the walker to the right (left) neighboring site if the state is $\ket{H}$ ($\ket{V}$), respectively. Each step of the QW consists of the following sequence of operations: $C_{\theta_2 /2}$, $G_\gamma$, $S$, $C_{\theta_1}$, $S$, $G_\gamma$, and $C_{\theta_2 /2}$. This sequence is repeated at every step to realize the full nonunitary evolution. These dynamics can equivalently be described by an effective Hamiltonian $H_{\mathrm{eff}}$, with the Floquet operator $U=e^{-i H_{\mathrm{eff}}}$ \cite{PhysRevB.88.121406}. The quasienergy $E$ of $H_{\mathrm{eff}}$ is defined as $E=i\ln\lambda_E $, where $\lambda_E $ denotes the eigenvalue of $U$. The QW defined above can be realized in quantum optical systems, where the walker is shifted more than twice at each time step \cite{PhysRevA.98.063847}.
	
	Distinct from the systems described in Refs. \cite{xiao2017observation,PhysRevA.102.062202}, where the nonunitary components exhibit parity and time-reversal symmetry and support standard edge modes, our system features sublattice symmetry in the Floquet operator $U$. This symmetry is characterized by the relation $\Gamma U \Gamma^{-1} = U^{-1}$, where the sublattice symmetry operator is given by $\Gamma = \sum_n \ket{n} \bra{n} \otimes \sigma_x$, with $\sigma_x$ being the standard Pauli matrix. The presence of sublattice symmetry intrinsically protects the topological properties of the quasienergy spectrum. Moreover, the Floquet operator $U$ also respects time-reversal symmetry, $T U^* T^{-1}=U^{-1}$, and particle-hole symmetry, $\Xi U^* \Xi^{-1}=U$, where the time-reversal operator is $T = \sum_n \ket{n} \bra{n} \otimes \sigma_x $, and the particle-hole operator is $\Xi = \sum_n \ket{n} \bra{n} \otimes \sigma_0$ with $\sigma_0$ being a $2\times 2$ identity matrix. Accordingly, in the unitary limit, the system belongs to the $\text{BDI}$ class within the the Altland-Zirnbauer (AZ) classification \cite{PhysRevX.9.041015}, while for $\gamma\ne 0$, it falls into the $\text{D}^{\dagger}$ class in the extended $\text{AZ}^{\dagger}$ classification \cite{PhysRevX.9.041015}. Due to the nonunitary nature of our system combined with sublattice symmetry, we observe anomalous edge modes that exhibit a transfer phenomenon, which distinguish them from the standard edge modes reported in previous studies.
	
	For a QW with sublattice symmetry hosting a time-periodic feature, the quasienergy is $2\pi$ periodic \cite{PhysRevX.3.031005,Kitagawa2012j}, leading to an additional quasienergy gap at the zone boundary ($E=\pi$), alongside the conventional gap at $E=0$. Consequently, two distinct types of topological edge states may emerge, associated with nontrivial quasienergies at both $E=0$ and $E=\pi$ \cite{PhysRevA.82.033429,PhysRevB.88.121406}. To characterize the topological phase transitions in our system, we apply the winding number $\nu$, defined in quasimomentum $k$ space under PBCs via the global Berry phase: $\nu=\varphi_B/2\pi$, where $\varphi_B=\varphi_{Z+}+\varphi_{Z-}$, and
	\begin{equation}
		\varphi_{Z\pm}=-\oint dk\frac{\bra{\psi_{\pm}^{(L)}} i\nabla_k\ket{\psi_{\pm}^{(R)}}}{\braket{\psi_{\pm}^{(L)}}{\psi_{\pm}^{(R)}}}\label{eq5}.
	\end{equation}
	Here, the right eigenstates $\ket{\psi_{\pm}^{(R)}}$ and the left eigenstates $\ket{\psi_{\pm}^{(L)}}$ are defined by $U\ket{\psi_{\pm}^{(R)}}=e^{-iE_{\pm}(k)}\ket{\psi_{\pm}^{(R)}}$ and $U^{\dagger}\ket{\psi_{\pm}^{(L)}}=e^{iE^{*}_{\pm}(k)}\ket{\psi_{\pm}^{(L)}}$, with $E_{\pm}(k)=\pm\arccos D(k)$ and $D(k)=\cos\left[2\left(k+i\gamma\right)\right]\cos\theta_{1}\cos\theta_{2}- \sin\theta_{1}\sin\theta_{2}$. When $\gamma=0$, $\ket{\psi_{\pm}^{(R)}}=\ket{\psi_{\pm}^{(L)}}$. Following Ref. \cite{PhysRevB.88.121406}, we further introduce an alternative winding number $\nu^{\prime}$, defined from the global Berry phase of the Floquet operator in a different time frame, $U^{\prime}=C_{\theta _{1}/2}S G_{\gamma} C_{\theta _{2}} G_{\gamma} S C_{\theta _{1}/2}$. The bulk topological invariants $\left(\nu_0, \nu_\pi\right)$ can then be constructed from winding numbers $\nu$ and $\nu^{\prime}$ as 
	\begin{equation}
		\left(\nu_0, \nu_\pi\right)=\left(\frac{\nu+\nu^{\prime}}{2},\frac{\nu-\nu^{\prime}}{2}\right),
		\label{eq6} 
	\end{equation}
	which correspond to the emergence of quasienergy modes at $E=0$ and $E=\pi$, respectively.
	
	In the following, we compare the topological phase diagrams for both unitary and nonunitary cases, and demonstrate the the transfer characteristics of the edge states in the nonunitary QWs. 
	
	\section{TOPOLOGICAL PHASE TRANSITIONS FOR UNITARY QWs}
	
	\begin{figure*}
		\centering
		\includegraphics[width=\linewidth]{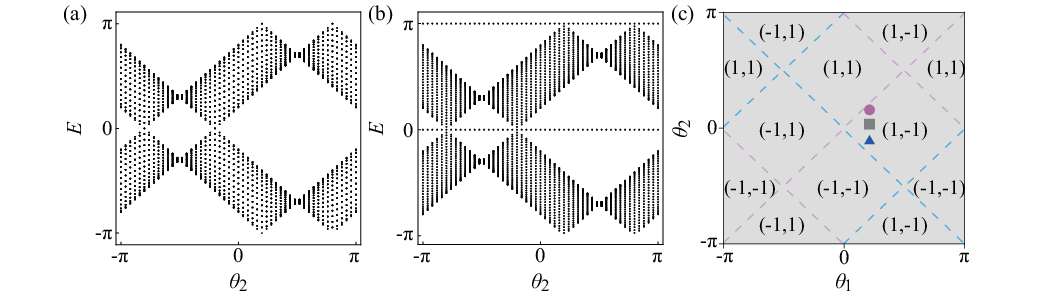}
		\caption{Floquet quasienergy spectra as functions of $\theta_2$ for the unitary QWs with $\theta_1=0.2\pi$ and $\gamma=0$ under (a) PBCs and (b) FBCs ($N=60$). (c) Topological phase diagram in the coin parameter space $\left( \theta_1, \theta_2 \right)$. The distinct phases are labeled by the winding numbers $\left(\nu_0,\nu_\pi\right)$, and the topological phase boundaries for $E = 0$ and $E = \pi$ are marked by blue and purple dashed lines, respectively. The blue triangle, gray rectangle, and purple circle denote parameter sets with $(\theta_{1},\theta_{2})=(0.2\pi,-0.15\pi)$, $(0.2\pi,0.05\pi)$, and $(0.2\pi,0.15\pi)$, respectively.}\label{fig1}
	\end{figure*}
	
	\begin{figure}
		\centering
		\includegraphics[width=\linewidth]{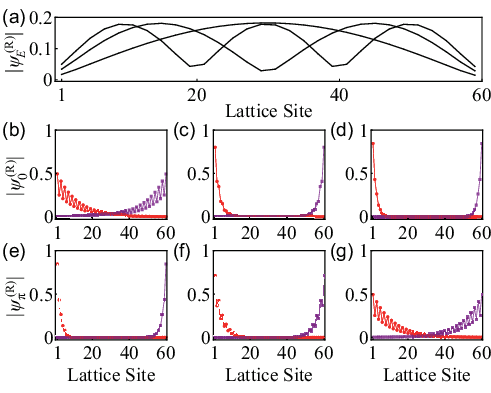}
		\caption{\label{fig2} 
			Spatial profiles of right eigenmodes in unitary QWs with $\gamma = 0$: (a) three arbitrarily chosen bulk states at $\left( \theta_1, \theta_2 \right) = (0.2\pi, -0.15\pi)$; (b)–(d) two degenerate $0$-modes; and (e)–(g) two degenerate $\pi$-modes. In (b)–(d) and (e)–(g) with $\theta_{1}=0.2\pi$, the columns from left to right correspond to $\theta_2=-0.15\pi$, $0.05\pi$, and $0.15\pi$, marked by the blue triangle, gray rectangle, and purple circle in Fig. \ref{fig1}(c), respectively. }
	\end{figure} 
	
	In this section, we investigate the topological phase transitions and the behavior of topological edge modes for QWs in the unitary limit($\gamma = 0$). This analysis leverages the well-established BBC in unitary systems, which asserts the consistency between bulk topological properties under PBCs and the emergence of edge states under FBCs. As a representative example, we set the coin parameter $\theta_1=0.2\pi$. Figures \ref{fig1}(a) and \ref{fig1}(b) display the Floquet quasienergy spectra as functions of $\theta_2$ under PBCs and FBCs, respectively. Under PBCs, the spectrum features well-defined topological gaps near $E=0$ and $E=\pm \pi$, which persist throughout the parameter range $\theta_{2}\in[-\pi,\pi]$, except at several discrete critical points. Specifically, the two spectral bands close at $\theta_{2}=-0.2\pi$ and $-0.8\pi$ for $E=0$, and at $\theta_{2}=0.2\pi$ and $0.8\pi$ for $E=\pi$. These topological features are preserved under FBCs with a system size of $N=60$, as shown in Fig. \ref{fig1}(b). Two distinct types of edge states emerge, corresponding to quasienergies $E=0$ and $E=\pi$, respectively, reflecting the conventional BBC in unitary systems. The number of $\tilde{\alpha}$ modes with $\tilde{\alpha}=\{0,\pi\}$ is determined by the absolute value of topological winding numbers $2\nu_{\alpha}$ defined in Eq \eqref{eq6}. Figure \ref{fig1}(c) shows the topological phase diagram in the coin parameter space $\left(\theta_1, \theta_2\right)$ for $\gamma=0$, where different phases are characterized by distinct winding numbers $\left(\nu_0, \nu_\pi\right)$. For arbitrary parameter choices, the winding numbers are nonzero except at gapless points, which signal topological phase transitions. These transitions occur either at $E = 0$ [blue dashed lines in Fig. \ref{fig1}(c)] or at $E = \pm \pi$ [purple dashed lines in Fig. \ref{fig1}(c)]. To further illustrate the behavior of topological edge states, we select several representative sets of $(\theta_1, \theta_2)$. For comparison, Fig. \ref{fig2}(a) shows the spatial profiles of three arbitrarily chosen right bulk states $|\psi^{(R)}_E|$ under FBCs for $(\theta_1,\theta_2)=(0.2\pi,-0.15\pi)$, clearly displaying the extended nature of bulk states. The spatial profiles $|\psi^{(R)}_{\tilde{\alpha}}|$ of the two degenerate edge-localized zero modes and $\pi$ modes are presented in Figs. \ref{fig2}(b) to \ref{fig2}(d) and Figs. \ref{fig2}(e) to \ref{fig2}(g), respectively, for $\theta_1=0.2\pi$. From left to right, these figures correspond to $\theta_2=-0.15\pi$, $0.05\pi$, and $0.15\pi$, which are marked by the blue triangle, gray rectangle, and purple circle in Fig. \ref{fig1}(c), respectively. Notably, the localization strengths of the edge modes vary with the coin parameters $\theta_2$.
	
	\begin{figure}
		\centering
		\includegraphics[width=\linewidth]{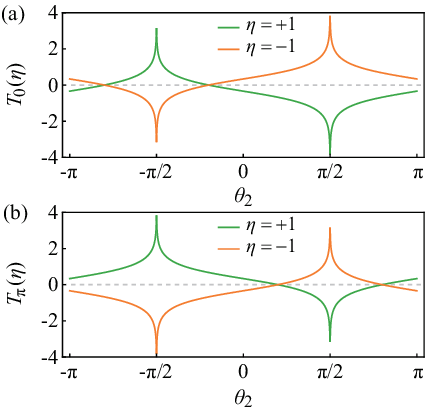}
		\caption{\label{fig3} 
			$T_{\tilde{\alpha}}(\eta)$ of the $\tilde{\alpha}$ modes as functions of $\theta_2$ with $\theta_{1}=0.2\pi$ for (a) $\tilde{\alpha}=0$ and (b) $\tilde{\alpha}=\pi$.}
	\end{figure} 
	
	\begin{figure*}
	\centering
	\includegraphics[width=\linewidth]{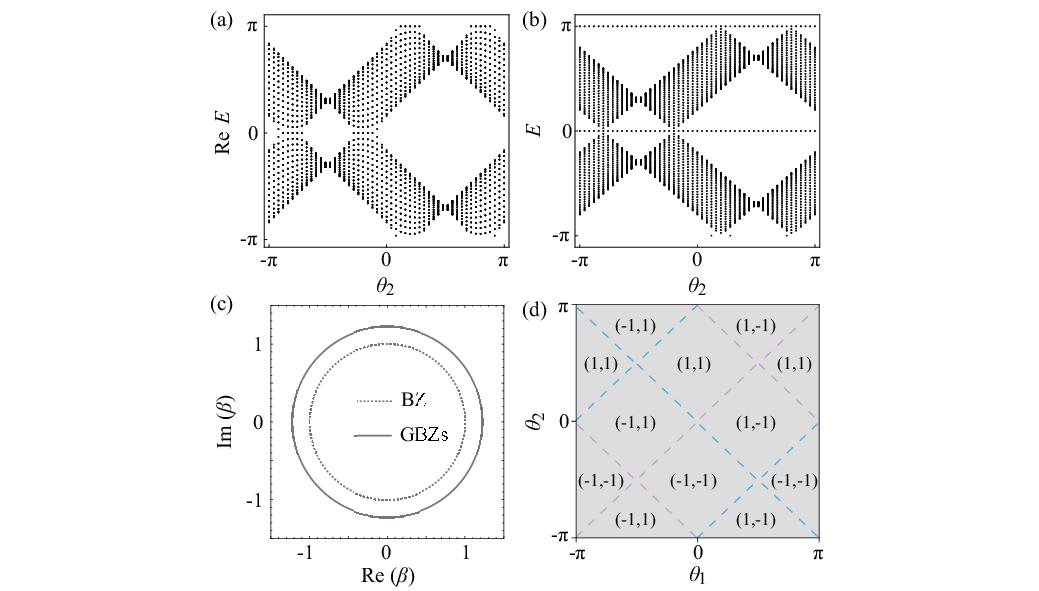}
	\caption{\label{fig4} 
		Floquet quasienergy spectra as functions of $\theta_2$ for the nonunitary QWs with $\theta_1=0.2\pi$ and $\gamma=0.2$ under (a) PBCs and (b) FBCs ($N=60$). (c) GBZs parametrized by $\beta$ in the complex plane. (d) Non-Bloch topological phase diagram. The phases are labeled by the non-Bloch winding numbers $\left(\tilde{\nu}_0,\tilde{\nu}_\pi\right)$. The topological phase boundaries for non-Bloch quasienergy $0$ and $\pi$ are marked by blue and purple dashed lines, respectively.}
	\end{figure*}		
	
	We employ the inverse of localization length to characterize the localization properties of the edge modes, which has been widely used to investigate localization phenomena in disordered systems \cite{PhysRevLett.47.1546,PhysRevLett.125.073204}, as well as the emergence of edge states in the field of topological insulators \cite{PhysRevLett.113.046802,PhysRevA.105.063327}. The right eigenstates with quasienergy $E=\tilde{\alpha}$ can be expressed as
	\begin{equation}
		\ket{\psi_{\tilde{\alpha}}^{(R)}}=\sum_{n} a_{\tilde{\alpha},n}\ket{n}\ket{H}+b_{\tilde{\alpha},n}\ket{n}\ket{V}.
		\label{eq7}
	\end{equation}
	Due to the sublattice symmetry of the Floquet operator $U$, the $\tilde{\alpha}$ modes are also eigenstates of the sublattice operator, satisfying $\Gamma \ket{\psi_{\tilde{\alpha}}^{(R)}}=\eta \ket{\psi_{\tilde{\alpha}}^{(R)}}$ with $\eta=\pm 1$. For $\eta=1$, we have $a_{\tilde{\alpha},n}=b_{\tilde{\alpha},n}$, whereas for $\eta=-1$, $a_{\tilde{\alpha},n}=-b_{\tilde{\alpha},n}$. Accordingly, the right eigenstates of the $\tilde{\alpha}$-modes can be written as (see the Appendix for details):
	\begin{equation}
		\ket{\psi_{\tilde{\alpha},\eta}^{(R)}}= \sum_{m=1}^{N/2}{\left[a_{\tilde{\alpha},2m-1}^{(\eta)}\ket{2m-1}+a_{\tilde{\alpha},2m}^{(\eta)}\ket{2m}\right]} \otimes \ket{C_{\eta}},
		\label{eq8} 
	\end{equation}
	where $\ket{C_{\eta}}=\ket{H}+\eta\ket{V}$, and 
	\begin{equation}
		\begin{cases}
			a_{\tilde{\alpha},2m-1}^{(\eta)}=e^{2(m-1) T_{\tilde{\alpha}}(\eta)}\sqrt{\frac{1-e^{4 T_{\tilde{\alpha}}(\eta)}}{2(1+Z^{2\eta}_{\tilde{\alpha}})[1-e^{2 N  T_{\tilde{\alpha}}(\eta)}]}},\\
			a_{\tilde{\alpha},2m}^{(\eta)}=\eta Z_{\tilde{\alpha}}^{\eta}\alpha_{\tilde{\alpha},2m-1}^{(\eta)},
		\end{cases}
		\label{eq9} 
	\end{equation}
	with the lattice size $N$ assumed to be even without loss of generality. Here, the absolute value of $T_{\tilde{\alpha}}(\eta)$, denoted as $|T_{\tilde{\alpha}}(\eta)|$, quantifies the inverse of the localization length for the $\tilde{\alpha}$ modes, where
	\begin{equation}
		T_{\tilde{\alpha}}(\eta)=\frac{1}{2}\eta\ln \left|Z_{\tilde{\alpha}} R_{2}\right|,
		\label{eq14} 
	\end{equation}
	with $Z_{0}=R_{1}$, $Z_{\pi}=-R_{1}^{-1}$, and  $R_{\tilde{\beta}=\{1,2\}}=\sec\theta_{\tilde{\beta}}-\tan\theta_{\tilde{\beta}}$. Edge modes at $E=\tilde{\alpha}$ emerge when $|T_{\tilde{\alpha}}(\eta)|>0$. The topological phase transitions are identified by the condition $|T_{\tilde{\alpha}}(\eta)|=0$ in Eq. \eqref{eq14}, leading to the following criteria 
	\begin{equation}
		\begin{cases}
			\theta_{2}=-\theta_{1}~\text{or}~\theta_{2}=\pm\pi+\theta_{1}&\text{for}~E=0\\
			\theta_{2}=\theta_{1}~\text{or}~\theta_{2}=\pm\pi-\theta_{1}&\text{for}~E=\pi
			\label{eq15} 
		\end{cases}.
	\end{equation}
	These analytical results are consistent with the numerical determined topological transition points shown in Fig. \ref{fig1}(c). 
	
	Figures \ref{fig3}(a) and \ref{fig3}(b) show $T_{0}(\eta)$ and $T_{\pi}(\eta)$ as functions of $\theta_2$ with $\theta_1=0.2\pi$, respectively. For the $E=0$ modes, when $\theta_2=-\pi$, the $0$-mode with $\eta=+1$ ($\eta=-1$) is localized at the left (right) boundary, corresponding to $T_{0}(\eta)<0$ ($>0$). As $\theta_2$ increases beyond $\theta_2=-0.8\pi$, where a topological phase transition occurs, $T_{0}(+1)$ and $T_{0}(-1)$ exchange their signs, indicating that the localized directions of the zero modes with different sublattice symmetries are inverted within $\theta_2 \in \left(-0.8\pi, -0.2\pi\right)$. With a further increase of $\theta_2$ past $-0.2\pi$, $T_{0}(+1)$ and $T_{0}(-1)$ exchange their signs again, so the $0$-mode with $\eta=-1$ becomes right-localized, while the one with $\eta=+1$ returns to being left-localized. As $\theta_2$ approaches these transition points, the magnitude of $\left|T_{0}(\eta)\right|$ decreases, indicating that both zero modes $\eta=\pm 1$ become less localized and penetrate deeper into the bulk. Similarly, for the $E=\pi$ modes with $\theta_1=0.2\pi$, phase transition points are localized at $\theta_2=0.2\pi$ and $0.8\pi$, where $T_{\pi}(\eta)=0$. When $\theta_2$ crosses these topological phase transition points, $T_{\pi}(\eta)$ for different $\eta$ exchanges sign, indicating a reversal in the localized direction of the corresponding $\pi$ modes. As $\theta_2$ approaches these points, $\left| T_{\pi}(\eta)\right|$ also decreases, showing that both $\pi$ modes become increasingly delocalized and extend further into the bulk, mirroring the behavior observed for the zero modes. These results demonstrate that, for the unitary case, the sign changes and magnitude reduction of $T_{\tilde{\alpha}}(\eta)$ serve as clear signatures of topological phase transitions. Furthermore, the edge modes with opposite sublattice symmetries ($\eta=\pm 1$) are always localized at opposite boundaries, and the direction of localization reverses when the sign of $T_{\tilde{\alpha}}(\eta)$ changes.

	\section{TOPOLOGICAL PHASE TRANSITIONS AND TRANSFER OF EDGE STATES FOR NONUNITARY QWs}
	
	In this section, we turn to the nonunitary case by introducing a finite gain and loss parameter $\gamma$. As a representative example, we set $\gamma=0.2$ and $\theta_1=0.2\pi$, and analyze the topological properties under PBCs and FBCs. Figure \ref{fig4}(a) shows the real part of the quasienergy spectrum, $\mathrm{Re}(E)$, as a function of $\theta_2$ under PBCs. Unlike the unitary case [see Fig. \ref{fig1}(a)], where the spectral bands close only at discrete points, here the two spectral bands of the real part form closed line segments at $\mathrm{Re}(E)=0$ and $\mathrm{Re}(E)=\pi$. In contrast, under FBCs, Fig. \ref{fig4}(b) presents the quasienergy spectrum $E$ as a function of $\theta_2$. In this case, the two spectral bands close at discrete critical points, resembling the behavior found in the unitary regime [see Fig. \ref{fig1}(b)]. These differences underscore the breakdown of the conventional BBC in the nonunitary regime.
	
	\begin{figure}
		\centering
		\includegraphics[width=\columnwidth]{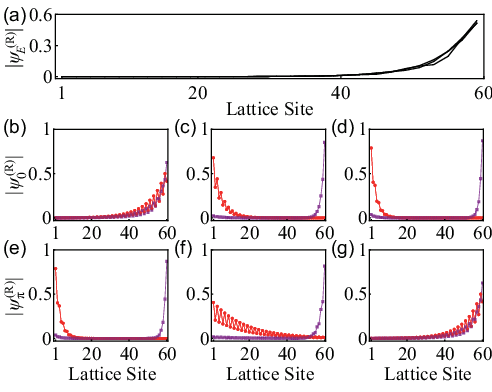}
		\caption{\label{fig5} 
			Spatial profiles of right eigenmodes in nonunitary QWs with $\gamma = 0.2$: (a) three arbitrarily chosen bulk states; (b)–(d) two degenerate zero modes; and (e)–(g) two degenerate $\pi$ modes. The coin parameters $\left( \theta_1, \theta_2 \right)$ are the same as those used in Fig. \ref{fig2}.}
	\end{figure}  
	
	\begin{figure}
		\centering
		\includegraphics[width=\linewidth]{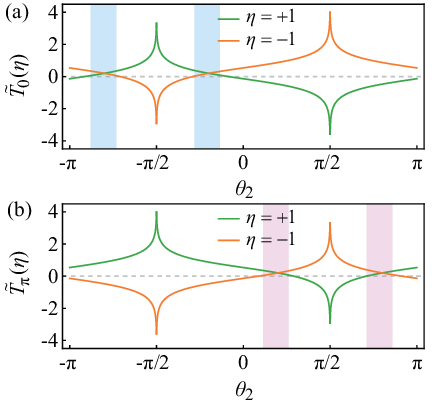}
		\caption{\label{fig5-5} 
			$\tilde{T}_{\tilde{\alpha}}(\eta)$ of the $\tilde{\alpha}$ modes as functions of $\theta_2$ for (a) $\tilde{\alpha}=0$ and (b) $\tilde{\alpha}=\pi$ in nonunitary QWs with $\gamma=0.2$ and $\theta_{1}=0.2\pi$. Regions with both $\tilde{T}_{0}(+1)>0$ and $\tilde{T}_{0}(-1)>0$ are shaded blue in (a), and regions with both $\tilde{T}_{\pi}(+1)>0$ and $\tilde{T}_{\pi}(-1)>0$ are shaded purple in (b).}
	\end{figure} 
	
	To restore BBC, one can employ the non-Bloch band theory \cite{PhysRevLett.121.086803,PhysRevLett.123.066404} to calculate the non-Bloch  winding numbers in Eq. \eqref{eq6} via contour integration along the generalized Brillouin zone (GBZ), rather than the standard Brillouin zone. To construct the GBZ, the standard Bloch phase factor $e^{ik}$ is replaced by a complex parameter $\beta$. Following the method proposed by Yokomizo \textit{et al}. for determining the continuum bands \cite{PhysRevLett.121.086803,PhysRevLett.123.066404,PhysRevLett.125.186802}, the trajectory of the GBZ, given by $\beta=e^{\gamma}e^{ip}$, can be obtained for our nonunitary QWs in the complex plane by varying the modified quasi-momentum $p$ from $0$ to $2\pi$, as illustrated in Fig. \ref{fig4}(c) with $\gamma=0.2$ as an example. By performing the integration along the GBZ, the non-Bloch winding numbers $\left(\tilde{\nu}_0,\tilde{\nu}_\pi\right)$ are evaluated in $\theta_{1}$-$\theta_{2}$ plane, as demonstrated in Fig. \ref{fig4}(d). Remarkably, it is observed that the resulting non-Bloch topological phase diagram is identical to its unitary counterpart [see Fig. \ref{fig1}(c)]. 
	
	The topological invariants $\tilde{\nu}_{\tilde{\alpha}}$ effectively predict the emergence of edge states at $E={\tilde{\alpha}}$ under FBCs, as demonstrated in Fig. \ref{fig4}(b). Figure \ref{fig5} presents the right eigenmode profiles under FBCs for the nonunitary case, with the coin parameters $\left(\theta_1, \theta_2\right)$ identical to those used in Fig. \ref{fig2}. Specifically, Fig. \ref{fig5}(a) displays the profiles of three random selected right bulk modes $|\psi^{(R)}_{E}|$, which clearly exhibit localization toward the right boundary, a hallmark of the NHSE. This phenomenon can be attributed to the GBZ, which forms a circular trajectory in the complex plane with a radius $e^{\gamma}>1$, corresponding to a right-directed skin effect. However, unlike the unitary case where degenerate edge states are localized at opposite boundaries with the identical localization length, the topologically protected edge states in the nonunitary case exhibit an anomalous spatial transfer. This behavior is shown in Figs. \ref{fig5}(b) to \ref{fig5}(d) and Figs. \ref{fig5}(e) to \ref{fig5}(g), where $\theta_1=0.2\pi$ corresponds to the profiles of the zero modes $|\psi^{(R)}_{0}|$ and the $\pi$ modes $|\psi^{(R)}_{\pi}|$, respectively. For the double-degenerate $E=0$ edge modes, when $\theta_2=0.15\pi$, which is far from the zero mode phase transition boundary [see in Fig. \ref{fig5}(d)], both edge modes are localized at different boundaries, similar to the unitary case. However, the localization strength of the left-localized zero mode is slightly weaker than that of the right-localized one. As $\theta_2$ decreases toward the phase transition point $\theta_2=-0.2\pi$, the inverse of localization strength of the left-localized zero mode gradually diminishes, while the right-localized modes remains largely unchanged, as shown in Fig. \ref{fig5}(c) for $\theta_2=0.05\pi$. When $\theta_2=-0.15\pi$ approaches the phase transition point [see in Fig. \ref{fig5}(b)], both zero modes become right-localized. Similarly, the doubly degenerate $E=\pi$ edge modes display analogous transfer behavior as $\left(\theta_{1},\theta_{2}\right)$ approach the $E=\pi$ phase transition boundaries, as depicted in Figs. \ref{fig5}(e) to \ref{fig5}(g).	
	
	\begin{figure*}
		\centering
		\includegraphics[width=\linewidth]{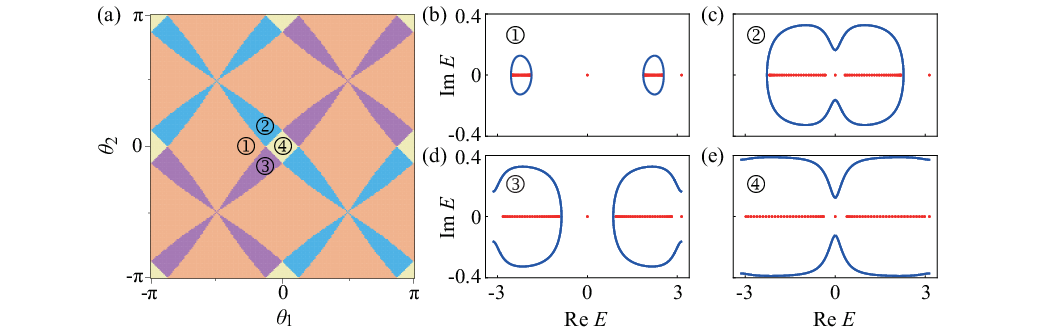}
		\caption{\label{fig6} 
			(a) Transition diagram of edge states in the coin parameter space $\left( \theta_1, \theta_2 \right)$ for the nonunitary QWs with $\gamma = 0.2$. In region \ding{172}, the degenerate zero modes and $\pi$ modes are localized at the opposite boundaries. In region \ding{173}, two zero modes are localized at the same boundary, while two $\pi$ modes remain at opposite boundaries. In region \ding{174}, two $\pi$ modes are localized at the same boundary, while two zero modes remain at opposite boundaries. In region \ding{175}, both zero modes and $\pi$ modes are localized at the same boundary. (b)–(e) Quaienergy spectra in the complex energy plane, with the corresponding parameter sets $\left( \theta_1, \theta_2 \right) =\left( -0.4\pi, -0.2\pi \right)$, $\left( -0.2\pi, 0.1\pi \right)$, $\left( -0.2\pi, 0.1\pi \right)$, and $\left( 0.04\pi,0.08\pi \right)$, respectively. The blue and red dots represent the spectra for PBCs and FBCs, respectively.}
	\end{figure*} 	
	
	To analyze the transfer behavior of the edge states, we employ the inverse of localization length. We analytically derive the right eigenstates $\ket{\psi_{\tilde{\alpha},\eta}^{(R)}}$ with $E=\tilde{\alpha}$ for the nonunitary system, expressed similarly to Eq. (\ref{eq8}). The amplitudes of the edge modes at odd and even lattice sites are given respectively by (see the Appendix for details):
	\begin{equation}
		\begin{cases}
			a_{\tilde{\alpha},2m-1}^{(\eta)}=e^{2(m-1)\tilde{T}_{\tilde{\alpha}}(\eta)}\sqrt{\frac{1-e^{4\tilde{T}_{\tilde{\alpha}}(\eta)}}{2(1+Z^{2\eta}_{\tilde{\alpha}})[1-e^{2 N \tilde{T}_{\tilde{\alpha}}(\eta)}]}},\\
			a_{\tilde{\alpha},2m}^{(\eta)}=\eta Z_{\tilde{\alpha}}^{\eta}\alpha_{\tilde{\alpha},2m-1}^{(\eta)},
		\end{cases}
		\label{eq12} 
	\end{equation}
	where $\tilde{T}_{\tilde{\alpha}}(\eta)=\gamma+T_{\tilde{\alpha}}(\eta)$ represents the combined contribution of the gain and loss parameter $\gamma$ and $T_{\tilde{\alpha}}(\eta)$. The transfer behavior of the edge states arises from the interplay between nonunitarity and topology. Figures \ref{fig5-5}(a) and \ref{fig5-5}(b) show $\tilde{T}_{0}(\eta)$ and $\tilde{T}_{\pi}(\eta)$ as functions of $\theta_2$ with $\gamma=0.2$ and $\theta_1=0.2\pi$, respectively. For the $E=0$ modes, when $\theta_2=-\pi$, $\tilde{T}_{0}(+1)<0$ ($\tilde{T}_{0}(-1)>0$), indicating that the zero mode with $\eta=+1$ ($\eta=-1$) is localized at the left (right) boundary. As $\theta_2$ increases to $-0.9138\pi$, $\tilde{T}_{0}(+1)$ increases to zero, while $\tilde{T}_{0}(-1)$ decreases but remains positive. In the region $\theta_2\in(-0.9138\pi,-0.7095\pi)$, both $\tilde{T}_{0}(+1)$ and $\tilde{T}_{0}(-1)$ are positive, meaning both zero modes with different sublattice symmetries are localized at the right boundary. Notably, $\tilde{T}_{0}(+1)$ and $\tilde{T}_{0}(-1)$ intersect at $\tilde{T}_{0}(-1)$=$\tilde{T}_{0}(+1)$=$\gamma$, where both zero modes share the same localization length. As $\theta_2$ further increases into $(-0.7095\pi,-0.2905\pi)$, $\tilde{T}_{0}(+1)>0$ and $\tilde{T}_{0}(-1)<0$, so the zero mode with $\eta=+1$ ($\eta=-1$) is localized at the right (left) boundary. In the region $\theta_2\in (-0.2905\pi, -0.0862\pi)$, both $\tilde{T}_{0}(\eta)>0$, so both zero modes are again localized at the right boundary. When $\theta_2$ exceeds $-0.0862\pi$, $\tilde{T}_{0}(+1)<0$ and $\tilde{T}_{0}(-1)>0$, restoring the localization of the two zero modes to opposite boundaries. For the $\pi$ modes with $\gamma=0.2$ and $\theta_1=0.2\pi$, $\tilde{T}_{\pi}(\eta)$ exhibits similar behavior. In the regions $\theta_2 \in (0.0862\pi, 0.2905\pi)$ and $\theta_2 \in(0.7095\pi, 0.9138\pi)$, both $\tilde{T}_{\pi}(\eta)>0$, implying both $\pi$ modes with different sublattice symmetries are localized at the same boundary. In the other regions, the two $\pi$ modes are localized at opposite boundaries. As $\gamma$ increases, the regions in $\theta_2$ where both $\tilde{\alpha}$ modes are localized at the same boundary gradually expand, reflecting the growing influence of nonunitarity compared to topology. In the nonunitary case, nonunitarity modifies the characteristics of the edge modes, leading to the emergence of both $\tilde{\alpha}$ modes with different sublattice symmetries being localized the same boundary, a phenomenon absent in the unitary case.

	We define $\mathcal{S}(0)=\mathrm{sgn}{[\tilde{T}_{0}(+1)\tilde{T}_{0}(-1)]}$ and 
	$\mathcal{S}(\pi)=\mathrm{sgn}{[\tilde{T}_{\pi}(+1)\tilde{T}_{\pi}(-1)]}$ to characterize the transfer behavior of edge states. The corresponding transition diagram for $\gamma=0.2$ is shown in Fig. \ref{fig6}(a). In region \ding{172}, where $\mathcal{S}(0)=-1$ and $\mathcal{S}(\pi)=-1$, the degenerate zero modes and $\pi$ modes are localized at opposite boundaries. In region \ding{173} [$\mathcal{S}(0)=+1$ and $\mathcal{S}(\pi)=-1$], two degenerate zero modes are localized at the same boundary, while the $\pi$ modes are localized at the opposite boundaries. Conversely, in region \ding{174} [$\mathcal{S}(0)=-1$ and $\mathcal{S}(\pi)=+1$], the two degenerate $\pi$ modes are localized at the same boundary, with the zero modes at the opposite boundaries. Finally, in region \ding{175}, where $\mathcal{S}(0)=+1$ and $\mathcal{S}(\pi)=+1$, both zero modes and $\pi$ modes are localized at the same boundary. The sign structure of $\mathcal{S}(0)$ and $\mathcal{S}(\pi)$ provides a clear classification of edge state localization, revealing a rich variety of boundary behaviors that depend sensitively on the system parameters.
	
	These distinct transfer behaviors can be further elucidated by analyzing the relationship between the quasienergy spectra under PBCs and FBCs. As shown in Figs. \ref{fig6}(b) to \ref{fig6}(e), the blue and red dots represent the spectra for PBCs and FBCs, respectively. It is evident that the PBC spectra form intricate loops in the complex quaienergy plane, whereas the FBC spectra remain real. This contrast in spectral structures provides valuable insight into the localization and transfer behavior of topological edge modes. In Fig. \ref{fig6}(b), corresponding to region \ding{172}, both the $E=0$ and $E=\pi$ modes lie outside the spectral loops, indicating that the associated topological edge modes do not exhibit the transfer and remain localized at opposite boundaries. However, when the $E=0$ ($E=\pi$) modes are enclosed within the loop while the $E=\pi$ ($E=0$) modes remain outside, as shown in Fig. \ref{fig6}(c) [Fig. \ref{fig6}(d)], corresponding to the region \ding{173} (region \ding{174}), the zero modes ($\pi$ modes) undergo transfer and become localized at the same boundary, while the modes associated with the energy level outside the loop remain localized at opposite boundaries. Finally, as shown in Fig. \ref{fig6}(e) for region \ding{175}, when both the $E=0$ and $E=\pi$ modes are enclosed within the loop, both the zero and $\pi$ modes undergo transfer, resulting in all topological edge states being localized at the same boundary. Moreover, these transfer behaviors can also be characterized by the winding number for the $\tilde{\alpha}$ modes, defined as $W_{\tilde{\alpha}}=1/(2\pi i) \int_{0}^{2\pi}dk \partial_{k} \arg [H_{\mathrm{eff}}(k)-\tilde{\alpha}]$, where $\arg[\cdot]$ denotes the argument of a complex number \cite{PhysRevLett.125.126402,li2025size}. When $W_{\tilde{\alpha}}=0$, the degenerate $\tilde{\alpha}$ modes exhibit a standard distribution, localizing at opposite boundaries. In contrast, when $W_{\tilde{\alpha}}=1(-1)$, both degenerate $\tilde{\alpha}$ modes undergo transfer and become localized at the right (left) boundary. Our calculations show that the winding number for the $\tilde{\alpha}$ modes are consistent with the results presented in Fig. \ref{fig6}(a). Overall, the relationship between the spectral topological structure and the localization of edge modes provides a clear spectral signature for identifying transfer behaviors.
	
	\section{CONCLUSION}
	In summary, we systematically studied topological phase transitions and edge-state behaviors in a time-multiplexed nonunitary QW with sublattice symmetry. In the unitary regime, conventional BBC holds, with edge modes localized at opposite boundaries. In the nonunitary regime, non-Hermitian skin effects emerge, and the conventional correspondence breaks down. By employing non-Bloch band theory and generalized Brillouin zones, we restored a generalized BBC and revealed a unique transfer phenomenon: edge modes with different sublattice symmetries can localize at the same boundary. Furthermore, we established that the structure of spectral loops under PBCs offers a clear signature for these transfer behaviors. Our results deepen the understanding of nonunitary topological systems and provide guidance for future experimental and theoretical studies.
	
	\begin{acknowledgements}
	We are grateful to Linhu Li for helpful discussions. Z.X. is supported by the NSFC (Grants No. 12375016 and No. 12461160324) and Beijing National Laboratory for Condensed Matter Physics (Grant No. 2023BNLCMPKF001). This work was also supported by NSF for Shanxi Province (Grant No. 1331KSC).
	\end{acknowledgements}
	
	\subsection*{DATA AVAILABILITY}
	The data that support the findings of this article are not publicly available upon publication because it is not technically feasible and/or the cost of preparing, depositing, and hosting the data would be prohibitive within the terms of this research project. The data are available from the authors upon reasonable request.
	
	\appendix
	\renewcommand{\thesection}{}
	\begin{widetext}
		
		\section{ANALYTICAL SOLUTION OF EDGE STATES IN QWS}
		Here, we give a detailed solution to the topological edge states of the nonunitary QWs. For the Floquet operator $U$ defined in Eq. (\ref{eq1}), the right eigenstates with quasienergy $E=\tilde{\alpha}=\{0,\pi\}$ can be denoted as
		\begin{equation}\label{eq1a}
			\ket{\psi_{\tilde{\alpha}}^{(R)}}=\sum_{n} a_{\tilde{\alpha},n}\ket{n}\ket{H}+b_{\tilde{\alpha},n}\ket{n}\ket{V},
			\tag{A1}
		\end{equation}
		where $\ket{\psi_{\tilde{\alpha}}^{(R)}}$ is defined by $U\ket{\psi_{\tilde{\alpha}}^{(R)}}=\lambda_{\tilde{\alpha}}\ket{\psi_{\tilde{\alpha}}^{(R)}}$ with $\lambda_{0}=+1$ and $\lambda_{\pi}=-1$.
		From the eigenfunction of $U$, we have the equations of amplitudes as
		\begin{equation}\label{eq2a}
			\begin{cases}
				\lambda_{\tilde{\alpha}} a_{\tilde{\alpha},1} = \mathtt{S}_{1} b_{\tilde{\alpha},1} + \frac{1}{2} \mathtt{C}_{1} \left[ \mathtt{S}_{2} a_{\tilde{\alpha},2} + \left( 1+\mathtt{C}_{2} \right)b_{\tilde{\alpha},2} - e^{-2\gamma} \left( (1-\mathtt{C}_{2}) a_{\tilde{\alpha},3} + \mathtt{S}_{2} b_{\tilde{\alpha},3} \right) \right]\\
				
				\lambda_{\tilde{\alpha}} b_{\tilde{\alpha},1} = \mathtt{S}_{1} a_{\tilde{\alpha},1} + \frac{1}{2} \mathtt{C}_{1} \left[ \mathtt{S}_{2} b_{\tilde{\alpha},2} + \left( 1-\mathtt{C}_{2} \right) a_{\tilde{\alpha},2} + e^{-2\gamma} \left( (1+\mathtt{C}_{2}) b_{\tilde{\alpha},3} +\mathtt{S}_{2} a_{\tilde{\alpha},3} \right) \right]\\
				
				\lambda_{\tilde{\alpha}} a_{\tilde{\alpha},2} = \frac{1}{2} \mathtt{C}_{1} \left[ \mathtt{S}_{2} a_{\tilde{\alpha},1} + (1+\mathtt{C}_{2}) b_{\tilde{\alpha},1} - e^{-2\gamma} \left( (1-\mathtt{C}_{2}) a_{\tilde{\alpha},4} + \mathtt{S}_{2} b_{\tilde{\alpha},4} \right) \right] - \mathtt{S}_{1} \left( \mathtt{S}_{2} a_{\tilde{\alpha},2} + \mathtt{C}_{2} b_{\tilde{\alpha},2} \right) \\
				
				\lambda_{\tilde{\alpha}} b_{\tilde{\alpha},2} = \frac{1}{2} \mathtt{C}_{1} \left[ \mathtt{S}_{2} b_{\tilde{\alpha},1} + (1-\mathtt{C}_{2}) a_{\tilde{\alpha},1} + e^{-2\gamma} \left( (1+\mathtt{C}_{2}) b_{\tilde{\alpha},4} + \mathtt{S}_{2} a_{\tilde{\alpha},4} \right) \right] + \mathtt{S}_{1} \left( \mathtt{C}_{2} a_{\tilde{\alpha},2} - \mathtt{S}_{2} b_{\tilde{\alpha},2} \right) \\
				
				\lambda_{\tilde{\alpha}} a_{\tilde{\alpha},N-1} = \frac{1}{2} \mathtt{C}_{1} \left[ \left(1-\mathtt{C}_{2}\right) b_{\tilde{\alpha},N} - \mathtt{S}_{2} a_{\tilde{\alpha},N} + e^{2\gamma} \left( (1+\mathtt{C}_{2})a_{\tilde{\alpha},N-3} - \mathtt{S}_{2} b_{\tilde{\alpha},N-3} \right) \right] - \mathtt{S}_{1} \left( \mathtt{S}_{2} a_{\tilde{\alpha},N-1} + \mathtt{C}_{2} b_{\tilde{\alpha},N-1} \right) \\
				
				\lambda_{\tilde{\alpha}} b_{\tilde{\alpha},N-1} = \frac{1}{2} \mathtt{C}_{1} \left[ \left( 1+\mathtt{C}_{2} \right)a_{\tilde{\alpha},N} - \mathtt{S}_{2} b_{\tilde{\alpha},N} - e^{2\gamma} \left( (1-\mathtt{C}_{2})b_{\tilde{\alpha},N-3} - \mathtt{S}_{2} a_{\tilde{\alpha},N-3} \right)  \right] + \mathtt{S}_{1} \left( \mathtt{C}_{2} a_{\tilde{\alpha},N-1} - \mathtt{S}_{2} b_{\tilde{\alpha},N-1} \right) \\
				
				\lambda_{\tilde{\alpha}} a_{\tilde{\alpha},N} = -\mathtt{S}_{1} b_{\tilde{\alpha},N} + \frac{1}{2} \mathtt{C}_{1} \left[ \left(1-\mathtt{C}_{2} \right)b_{\tilde{\alpha},N-1} -\mathtt{S}_{2} a_{\tilde{\alpha},N-1} + e^{2\gamma} \left( (1+\mathtt{C}_{2})a_{\tilde{\alpha},N-2} - \mathtt{S}_{2} b_{\tilde{\alpha},N-2} \right)  \right] \\
				
				\lambda_{\tilde{\alpha}} b_{\tilde{\alpha},N} = -\mathtt{S}_{1} a_{\tilde{\alpha},N} + \frac{1}{2} \mathtt{C}_{1} \left[ \left(1+\mathtt{C}_{2}\right)a_{\tilde{\alpha},N-1} - \mathtt{S}_{2} b_{\tilde{\alpha},N-1} - e^{2\gamma} \left( (1-\mathtt{C}_{2}) b_{\tilde{\alpha},N-2}  - \mathtt{S}_{2} a_{\tilde{\alpha},N-2} \right) \right]
			\end{cases},
			\tag{A2}
		\end{equation}
		and 
		\begin{equation}\label{eq3a}
			\begin{cases}
				\lambda_{\tilde{\alpha}} a_{\tilde{\alpha},n} = \frac{1}{2} \mathtt{C}_{1} \left[ e^{2\gamma} \left( (\mathtt{C}_{2}+1)a_{\tilde{\alpha},n-2} - \mathtt{S}_{2} b_{n-2} \right)  + e^{-2\gamma} \left( \mathtt{S}_{2} b_{\tilde{\alpha},n+2}+(\mathtt{C}_{2}-1)a_{\tilde{\alpha},n+2}  \right) \right] - \mathtt{S}_{1} \left(\mathtt{S}_{2} a_{\tilde{\alpha},n} + \mathtt{C}_{2} b_{\tilde{\alpha},n} \right) \\
				
				\lambda_{\tilde{\alpha}} b_{\tilde{\alpha},n} = \frac{1}{2} \mathtt{C}_{1} \left[ e^{2\gamma} \left( (\mathtt{C}_{2}-1) b_{\tilde{\alpha},n-2}+\mathtt{S}_{2} a_{\tilde{\alpha},n-2} \right) + e^{-2\gamma} \left( \mathtt{S}_{2} a_{\tilde{\alpha},n+2} + (\mathtt{C}_{2}+1) b_{\tilde{\alpha},n+2} \right) \right] + \mathtt{S}_{1} \left( \mathtt{C}_{2} a_{\tilde{\alpha},n} - \mathtt{S}_{2} b_{\tilde{\alpha},n} \right) 
			\end{cases},
			\tag{A3}
		\end{equation}
		where $\mathtt{S}_{\tilde{\beta}=\left\{1,2\right\}} = \sin{\theta_{\tilde{\beta}}}$ and $\mathtt{C}_{\tilde{\beta}} = \cos{\theta_{\tilde{\beta}}}$.
		The Floquet operator $U$ exhibits sublattice symmetry with the relation $\Gamma U\Gamma={U}^{-1}$ and the symmetry operator $\Gamma=\sum_{n}{I_n\otimes\sigma_1}$. Edge states with quasienergy $E=\tilde{\alpha}$ are therefore also eigenstates of the sublattice operator, satisfying $\Gamma \ket{\psi_{\tilde{\alpha}}^{(R)}}=\eta \ket{\psi_{\tilde{\alpha}}^{(R)}}$ with $\eta=\pm 1$. For $\eta=+1$, we have $a_{\tilde{\alpha},n}=b_{\tilde{\alpha},n}$ whereas for $\eta=-1$, $a_{\tilde{\alpha},n}=-b_{\tilde{\alpha},n}$. Equivalently, the edge states are either in $\ket{+}=\ket{H}+\ket{V}$ or $\ket{-}=\ket{H}-\ket{V}$. 
		Specifically, edge states with quasienergy $E=\tilde{\alpha}$ are in $\ket{+}$, substituting $a_{\tilde{\alpha},n}=b_{\tilde{\alpha},n}$ into Eqs. (\ref{eq2a}) and (\ref{eq3a}), the equations of amplitudes reduce to the following
		\begin{equation}\label{eq6a}
			\begin{cases}
				a_{\tilde{\alpha},1}=e^{-2\gamma}Z_{\tilde{\alpha}}^{-1}R_2^{-1} a_{\tilde{\alpha},3}\\
				a_{\tilde{\alpha},2}=e^{-2\gamma}R_2^{-1} a_{\tilde{\alpha},3}\\
				a_{\tilde{\alpha},1}=e^{-2\gamma}Z_{\tilde{\alpha}}^{-2}R_2^{-1} a_{\tilde{\alpha},4}\\
				a_{\tilde{\alpha},2}=e^{-2\gamma}Z_{\tilde{\alpha}}^{-1}R_2^{-1} a_{\tilde{\alpha},4}\\
				a_{\tilde{\alpha},N-1}=e^{2\gamma}Z_{\tilde{\alpha}}R_2 a_{\tilde{\alpha},N-3}\\
				a_{\tilde{\alpha},N}=e^{2\gamma}Z_{\tilde{\alpha}}^2R_2 a_{\tilde{\alpha},N-3}\\
				a_{\tilde{\alpha},N-1}=e^{2\gamma}R_2 a_{\tilde{\alpha},N-2}\\
				a_{\tilde{\alpha},N}=e^{2\gamma}Z_{\tilde{\alpha}}R_2 a_{\tilde{\alpha},N-2}
			\end{cases},
			\tag{A4}
		\end{equation}
		and
		\begin{equation}\label{eq5a}
			\begin{cases}
				\left[\left(Z_{\tilde{\alpha}} R_2+Z_{\tilde{\alpha}}^{-1}R_2^{-1}\right)+\frac{1}{2}\prod_{\tilde{\beta}}\left(R_{\tilde{\beta}}^{-1}-R_{\tilde{\beta}}\right)\right] a_{\tilde{\alpha},n}=e^{2\gamma}\left(R_2+1\right) a_{\tilde{\alpha},n-2}-e^{-2\gamma}\left(R_2^{-1}-1\right) a_{\tilde{\alpha},n+2}\\
				
				\left[\left(Z_{\tilde{\alpha}} R_2^{-1}+Z_{\tilde{\alpha}}^{-1}R_2\right)+\frac{1}{2}\prod_{\tilde{\beta}}\left(R_{\tilde{\beta}}^{-1}-R_{\tilde{\beta}}\right)\right]a_{\tilde{\alpha},n}=-e^{2\gamma}\left(R_2-1\right)a_{\tilde{\alpha},n-2}+e^{-2\gamma}\left(R_2^{-1}+1\right)a_{\tilde{\alpha},n+2}
			\end{cases},
			\tag{A5}
		\end{equation}
		respectively. Here, $Z_{0}=R_1$, $Z_{\pi}=-R_{1}^{-1}$, and $R_{\tilde{\beta}}=\sec\theta_{\tilde{\beta}}-\tan\theta_{\tilde{\beta}}$. 
		By combining the two equations in Eq. (\ref{eq5a}), we obtain
		\begin{equation}\label{eq7a}
			\left(Z_{\tilde{\alpha}}R_2+\frac{1}{Z_{\tilde{\alpha}}R_2}\right) a_{\tilde{\alpha},n}=e^{2\gamma} a_{\tilde{\alpha},n-2}+e^{-2\gamma} a_{\tilde{\alpha},n+2}\ ,\ \ \ \ \ \text{for}\ \ \ 3\le n\le N-2.
			\tag{A6}
		\end{equation}
		When the lattice sites $n=3$ and $n=4$, the equation above the equation above takes the forms
		\begin{equation}\label{eq8a}
			\begin{cases}
				\left(Z_{\tilde{\alpha}}R_2+Z_{\tilde{\alpha}}^{-1}R_2^{-1}\right) a_{\tilde{\alpha},3}=e^{2\gamma} a_{\tilde{\alpha},1}+e^{-2\gamma} a_{\tilde{\alpha},5}\\
				\left(Z_{\tilde{\alpha}}R_2+Z_{\tilde{\alpha}}^{-1}R_2^{-1}\right) a_{\tilde{\alpha},4}=e^{2\gamma} a_{\tilde{\alpha},2}+e^{-2\gamma} a_{\tilde{\alpha},6}
			\end{cases},
			\tag{A7}
		\end{equation}
		respectively. Substituting both $ a_{\tilde{\alpha},1}=e^{-2\gamma}Z_{\tilde{\alpha}}^{-1}R_2^{-1} a_{\tilde{\alpha},3} $ and $ a_{\tilde{\alpha},2}=e^{-2\gamma}Z_{\tilde{\alpha}}^{-1}R_2^{-1} a_{\tilde{\alpha},4} $ from Eq. (\ref{eq6a}) into Eq. (\ref{eq8a}), the equations can be reformulated as
		\begin{equation}\label{eq9a}
			\begin{cases}
				Z_{\tilde{\alpha}}R_2 a_{\tilde{\alpha},3}=e^{-2\gamma} a_{\tilde{\alpha},5}\\
				Z_{\tilde{\alpha}}R_2 a_{\tilde{\alpha},4}=e^{-2\gamma} a_{\tilde{\alpha},6}\\
			\end{cases}.
			\tag{A8}
		\end{equation}
		By substituting  Eq. (\ref{eq9a}) into Eq. (\ref{eq7a})  and iteratively applying the procedure multiple times, this leads to
		\begin{equation}\label{eq10a}
			\begin{cases}
				Z_{\tilde{\alpha}}R_2 a_{\tilde{\alpha},5}=e^{-2\gamma} a_{\tilde{\alpha},7}\\
				Z_{\tilde{\alpha}}R_2 a_{\tilde{\alpha},7}=e^{-2\gamma} a_{\tilde{\alpha},9}\\
				~~~~~~~~~~~~\vdots\\
				Z_{\tilde{\alpha}}R_2 a_{\tilde{\alpha},2m-1}=e^{-2\gamma} a_{\tilde{\alpha},2m+1}\\
				~~~~~~~~~~~~\vdots\\
				Z_{\tilde{\alpha}}R_2 a_{\tilde{\alpha},N-3}=e^{-2\gamma} a_{\tilde{\alpha},N-1}\\
			\end{cases},
			\tag{A9}
		\end{equation}
		and
		\begin{equation}\label{eq11a}
			\begin{cases}
				Z_{\tilde{\alpha}}R_2 a_{\tilde{\alpha},6}=e^{-2\gamma} a_{\tilde{\alpha},8}\\
				Z_{\tilde{\alpha}}R_2 a_{\tilde{\alpha},8}=e^{-2\gamma} a_{\tilde{\alpha},10}\\
				~~~~~~~~~~~~\vdots\\
				Z_{\tilde{\alpha}}R_2 a_{\tilde{\alpha},2m}=e^{-2\gamma} a_{\tilde{\alpha},2m+2}\\
				~~~~~~~~~~~~\vdots\\
				Z_{\tilde{\alpha}}R_2 a_{\tilde{\alpha},N-2}=e^{-2\gamma} a_{\tilde{\alpha},N}\\
			\end{cases},
			\tag{A10}
		\end{equation}
		where system size $N$ is even without loss generally. From Eqs. (\ref{eq6a}) and (\ref{eq9a})-(\ref{eq11a}), the amplitudes at odd and even lattice sites are expressed in terms of $a_{\tilde{\alpha},1}$ and $a_{\tilde{\alpha},2}$, respectively, as
		\begin{equation}\label{eq12a}
			\begin{cases}
				a_{\tilde{\alpha},2m-1} = \left(e^{2\gamma}Z_{\tilde{\alpha}}R_2\right)^{m-1} a_{\tilde{\alpha},1}\\
				a_{\tilde{\alpha},2m} = \left(e^{2\gamma}Z_{\tilde{\alpha}}R_2\right)^{m-1} a_{\tilde{\alpha},2}
			\end{cases},
			\tag{A11}
		\end{equation}
		where $m=\left\{1,\ 2,\ 3,\ \ldots,\ \frac{N}{2}\right\}$. Combining Eq. (\ref{eq12a}) with $ a_{\tilde{\alpha},1}=e^{-2\gamma}Z_{\tilde{\alpha}}^{-1}R_2^{-1} a_{\tilde{\alpha},3} $ and $ a_{\tilde{\alpha},2}=e^{-2\gamma}R_2^{-1} a_{\tilde{\alpha},3}$ from Eq. (\ref{eq6a}), the relationship between the amplitudes at odd and even lattice sites can be expressed as
		\begin{equation}\label{eq13a}
			a_{\tilde{\alpha},2m}=Z_{\tilde{\alpha}} a_{\tilde{\alpha},2m-1}.
			\tag{A12}
		\end{equation}
		From the self-normalized condition $\left\langle\psi_{\tilde{\alpha}}^{(R)}\middle|\psi_{\tilde{\alpha}}^{(R)}\right\rangle=1$, it follows that
		\begin{equation}\label{eq14a}
			\sum_{n=1}^{N}\left| a_{\tilde{\alpha},n}\right|^2=\frac{1}{2}.
			\tag{A13}
		\end{equation}
		According to $ a_{\tilde{\alpha},2m}=Z_{\tilde{\alpha}} a_{\tilde{\alpha},2m-1}$ and $\left(e^{2\gamma}Z_{\tilde{\alpha}}R_2\right)^{m-1} a_{\tilde{\alpha},1}= a_{\tilde{\alpha},2m-1}$, Eq. (\ref{eq14a}) simplifies to
		\begin{equation}\label{eq16a}
			\left| a_{\tilde{\alpha},1}\right|^2\sum_{m=1}^{N/2}\left|e^{2\gamma}Z_{\tilde{\alpha}}R_2\right|^{2\left(m-1\right)}=\frac{1}{2\left(1+Z_{\tilde{\alpha}}^2\right)}.
			\tag{A14}
		\end{equation}
		When $\left|Z_{\tilde{\alpha}}R_2\right|=e^{-2\gamma}$, the equations above can be reduced to 
		\begin{equation}\label{eq17a}
			\left| a_{\tilde{\alpha},1}\right|=\sqrt{\frac{1}{N\left(1+Z_{\tilde{\alpha}}^2\right)}}.
			\tag{A15}
		\end{equation}
		When $\left|Z_{\tilde{\alpha}}R_2\right|\neq e^{-2\gamma}$, the Eq. (\ref{eq16a}) can be rewritten as
		\begin{equation}\label{eq18a}
			\left| a_{\tilde{\alpha},1}\right|=\sqrt{\frac{1-\left|e^{2\gamma}Z_{\tilde{\alpha}}R_2\right|^2}{2\left(1+Z_{\tilde{\alpha}}^2\right)\left(1-\left|e^{2\gamma}Z_{\tilde{\alpha}}R_2\right|^N\right)}}.
			\tag{A16}
		\end{equation}
		Thus, 
		\begin{equation}\label{eq19a}
			a_{\tilde{\alpha},2m-1}=
			\begin{cases}
				\sqrt{\frac{1}{N\left(1+Z_{\tilde{\alpha}}^2\right)}}&\text{for}~\left|Z_{\tilde{\alpha}}R_2\right|=e^{-2\gamma}\\
				\sqrt{\frac{1-\left|e^{2\gamma}Z_{\tilde{\alpha}}R_2\right|^2}{2\left(1+Z_{\tilde{\alpha}}^2\right)\left(1-\left|e^{2\gamma}Z_{\tilde{\alpha}}R_2\right|^N\right)}}\left|e^{2\gamma}Z_{\tilde{\alpha}}R_2\right|^{m-1}&\text{for}~\left|Z_{\tilde{\alpha}}R_2\right|\neq e^{-2\gamma}
			\end{cases},
			\tag{A17}
		\end{equation}
		and $ a_{\tilde{\alpha},2m}=Z_{\tilde{\alpha}} a_{\tilde{\alpha},2m-1}$ where $m=\left\{1,\ 2,\ 3,\ \ldots,\ \frac{N}{2}\right\}$. 
		
		By the same way, for edge states in $\ket{-}$, we derive
		\begin{equation}\label{eq20a}
			a_{\tilde{\alpha},2m-1}=
			\begin{cases}
				\sqrt{\frac{1}{N\left(1+Z_{1,E\ }^2\right)}}&\text{for}~\left|Z_{\tilde{\alpha}}R_2\right|=e^{-2\gamma}\\
				\sqrt{\frac{1-\left|e^{-2\gamma}Z_{\tilde{\alpha}}R_2\right|^{-2}}{2\left(1+Z_{\tilde{\alpha}}^{-2}\right)\left(1-\left|e^{-2\gamma}Z_{\tilde{\alpha}}R_2\right|^{-N}\right)}}\left|e^{-2\gamma}Z_{\tilde{\alpha}}R_2\right|^{-(m-1)}&\text{for}~\left|Z_{\tilde{\alpha}}R_2\right|\neq e^{-2\gamma}
			\end{cases},
			\tag{A18}
		\end{equation}
		and $ a_{\tilde{\alpha},2m}=-Z_{\tilde{\alpha}}^{-1} a_{\tilde{\alpha},2m-1}$ where $m=\left\{1,\ 2,\ 3,\ \ldots,\ \frac{N}{2}\right\}$. 
		
		In summary, the right eigenstates of the $\tilde{\alpha}$-modes can be written as:
		\begin{equation}\label{eq21a}
			\ket{\psi_{\tilde{\alpha},\eta}^{(R)}}= \sum_{m=1}^{N/2}{\left[a_{\tilde{\alpha},2m-1}^{(\eta)}\ket{2m-1}+a_{\tilde{\alpha},2m}^{(\eta)}\ket{2m}\right]} \otimes \ket{C_{\eta}},
			\tag{A19} 
		\end{equation}
		where $\ket{C_{\eta}}=\ket{H}+\eta\ket{V}$ and $\eta=\pm1$, and 
		\begin{equation}\label{eq22a}
			\begin{cases}
				a_{\tilde{\alpha},2m-1}^{(\eta)}=e^{2(m-1)\tilde{T}_{\tilde{\alpha}}(\eta)}\sqrt{\frac{1-e^{4\tilde{T}_{\tilde{\alpha}}(\eta)}}{2(1+Z^{2\eta}_{\tilde{\alpha}})[1-e^{2 N \tilde{T}_{\tilde{\alpha}}(\eta)}]}}\\
				a_{\tilde{\alpha},2m}^{(\eta)}=\eta Z_{\tilde{\alpha}}^{\eta}\alpha_{\tilde{\alpha},2m-1}^{(\eta)}
			\end{cases},
			\tag{A20} 
		\end{equation}	
		where the inverse of the localization length for the nonunitary QWs $\tilde{T}_{\tilde{\alpha}}(\eta)=\gamma+T_{\tilde{\alpha}}(\eta)$ depends on both the gain-loss parameter $\gamma$ and $T_{\tilde{\alpha}}(\eta)=\frac{1}{2}\eta\ln \left|Z_{\tilde{\alpha}} R_{2}\right|$.
		
	\end{widetext}

	\bibliographystyle{apsrev4-1}

\end{document}